\documentclass[aps,pra,twocolumn,floatfix]{revtex4}
\usepackage{epsfig}
\usepackage{graphicx}
\usepackage{dcolumn}
\usepackage{color}
\usepackage{longtable}
\usepackage{amsthm,amsmath}

\begin{document}

\title{Higher-component quadrupole polarizabilities: Estimations for the clock states of the alkaline earth-metal ions}

\author{Sukhjit Singh$^a$\footnote{Email: sukhjitsingh88@gmail.com}, Mandeep Kaur$^a$, Bindiya Arora$^a$\footnote{Email: bindiya.phy@gndu.ac.in} and B. K. Sahoo$^{b}$\footnote{Email: bijaya@prl.res.in} }
\affiliation{$^a$Department of Physics, Guru Nanak Dev University, Amritsar, Punjab-143005, India}
\affiliation{$^b$Atomic, Molecular and Optical Physics Division, Physical Research Laboratory, Navrangpura, Ahmedabad-380009, India}

\date{Received date; Accepted date}

\begin{abstract}
Derivations for the higher tensor components of the quadrupole polarizabilities are given and their values for the metastable 
states of the Ca$^+$, Sr$^+$ and Ba$^+$ alkaline earth-metal ions are estimated. We also give the scalar quadrupole 
polarizabilities of the ground and metastable states of these ions to compare our results with the previously available 
theoretical and experimental results. Reasonably good agreement between our calculations with the previous values of scalar 
quadrupole polarizabilities demonstrate their correctness. The reported scalar and tensor quadrupole polarizabilities 
could be very useful to estimate the uncertainties due to the gradient of the electric fields in the clock frequencies of 
the above alkaline earth-metal ions when accuracies of these frequency measurements attain below 10$^{-19}$ precision level.
\end{abstract}
\maketitle

\section{Introduction}

Last two decades have witnessed an enormous advancement in many high-precision measurements using singly charged ions 
such as parity violation \cite{bouchiat,fortson,wansbeek}, quantum phase transitions \cite{islam}, atomic clocks, quantum transport
of systems for a better understanding of fundamental physics etc. \cite{wood,grim,haffner,lukin}. The development has been 
attributed to the fact that ions can be manipulated favorably by employing optical cooling and trapping schemes. Oscillating 
electric fields due to lasers applied to cool and trap the atomic systems inevitably yield Stark shifts in the systems. Precise 
estimates of these energy shifts are prerequisite for the above experiments.

When an atomic system is exposed to an electric field, the charge inside the system redistributes itself. The modified distribution 
of charge can be expressed in terms of the electric multipole moments. The lowest-order electric moment referred to as the dipole (E1) moment, 
one more order higher multipole moment in the series known as the quadrupole (E2) moment and so on. Most of the times the second-order energy 
shifts due to the dipole moment, expressed in terms of the dipole polarizability ($\alpha^{E1})$, are estimated owing to their dominant 
contributions. For convenience, $\alpha^{E1}$ of a state is determined by expressing in terms of scalar, vector and tensor components. If the light is not linearly polarized 
the scalar and vector parts contribute to all the atomic states whereas the tensor component contributes to the states with total angular
momentum $j > 1/2$. Similarly, atomic levels interacting with the gradient of electric field can also give significant energy shifts in the 
atomic systems under consideration for high-precision experiments. Estimations of these shifts are important for accurate knowledge 
of total systematic shifts; especially in the atomic clock frequency measurements that are aiming to attain precisions below 10$^{-19}$ 
level. The first-order energy shifts of the states with the total angular momentum $j>1/2$ due to the gradient of the electric field can be 
estimated with the knowledge of electric quadrupole moments of these states. The second-order energy shifts of the atomic states due to the 
gradient of the electric field can be determined using the quadrupole polarizabilities ($\alpha^{E2}$) of the states. Likewise $\alpha^{E1}$, 
we can also conveniently express $\alpha^{E2}$ into scalar and high-order tensor components. Contributions from the scalar components are 
studied for many atomic systems \cite{derevianko,Lim10,arora1,arorab,sukhjit1,sukhjitnew,porsev1,blundell,safronova22,jasmeet,Iskrenova,Arora86,tang}, 
but contributions from the tensor components are often neglected. To our knowledge, expressions for these higher components are not deduced 
so far, which can give finite contributions to the states with $j > 1/2$.

For the ultra-high precision measurements, a detailed analysis of contributions from different components of $\alpha^{E2}$ would be required. 
Their knowledge is also essential in the evaluation of scattering properties of collisions between a neutral atom and its ions~\cite{cote}. 
They also provide information on the processes involving the collision of two neutral or charged species, where polarizability elucidates the
behavior of one in response of the other. Other important properties such as total energy shift, long range inter-atomic interactions, ion 
mobility, van der Waals constant between different systems etc. can be evaluated from the accurate values of the polarizabilities 
of the atoms and ions~\cite{bonin}. Moreover, finding magic wavelengths for the clock transitions in the neutral atoms and singly charged ions 
are now in high demand as they correspond to null differential polarizabilities and can offer frequency standards below 10$^{-19}$ precision 
level. Carrying out clock frequency measurements at the magic wavelengths is helpful to reduce uncertainty due to the Stark effects. Recently, 
these magic wavelengths were measured for the $S-D_{5/2}$ clock transition of the Ca$^+$ ion \cite{gao}. Following this measurement, a number of 
theoretical works have predicted magic wavelengths of the possible $S-D$ clocks transitions in other singly charged ions \cite{jasmeet,junjiang,atishroy}. From this point of view, contributions from the 
$\alpha^{E2}$ to the Stark shifts in the $S-D$ transitions of these singly charged ions should be investigated thoroughly.

In this work, we give first the derivations for the tensor components of $\alpha^{E2}$ and determine them for the metastable $D_{3/2}$ and $D_{5/2}$
states of the alkaline earth-metal Ca$^+$, Sr$^+$ and Ba$^+$ ions. We also determine the scalar components of $\alpha^{E2}$ values of 
these states along with the ground states of the above ions and compare them with the other available theoretical and experimental 
results to find out reliability of our calculations. 

\begin{table}[h!]
%\begin{ruledtabular}
%\begin{longtable}{lrrrr}
\caption{Contributions to the static scalar and rank 2 tensor component quadrupole polarizabilities (in a.u.) of the ground and $3D_{3/2}$ states of
Ca$^{+}$. The absolute values and uncertainties (denoted by $\delta$) are given from various contributions. The E2 matrix elements used for determining 
valence correlation contributions are also quoted.}\label{nsca}
\begin{ruledtabular}
\begin{tabular}{lrrrr}
\multicolumn{5}{c}{Ground state} \\
\multicolumn{1}{c}{Contribution} &
\multicolumn{1}{c}{$Q$} &
\multicolumn{1}{c}{$\delta Q$} &
\multicolumn{1}{c}{$\alpha_{n}^{E2(0)}$} &
\multicolumn{1}{c}{$\delta \alpha_{n}^{E2(0)}$}\\
\hline
 & & \\
\multicolumn{5}{l}{Main valence correlation} \\
4$S_{1/2}$  -  3$D_{3/2}$   &  7.971  &  0.064 &         204.32  &     3.28 \\
4$S_{1/2}$  -  4$D_{3/2}$   & 12.649  &  0.045 &         123.56  &     0.88 \\
4$S_{1/2}$  -  5$D_{3/2}$   &  4.217  &  0.021 &          10.73  &     0.11 \\
4$S_{1/2}$  -  6$D_{3/2}$   &  2.256  &  0.020 &           2.77  &     0.05 \\
4$S_{1/2}$  -  3$D_{5/2}$   &  9.785  &  0.081 &         306.53  &     5.08 \\
4$S_{1/2}$  -  4$D_{5/2}$   & 15.474  &  0.057 &         184.85  &     1.36 \\
4$S_{1/2}$  -  5$D_{5/2}$   &  5.168  &  0.026 &          16.12  &     0.16 \\
4$S_{1/2}$  -  6$D_{5/2}$   &  2.767  &  0.024 &           4.17  &     0.07 \\
& & \\
Tail($\alpha_{n,v}^{E2(k)}$)&&&           15  &       8 \\
$\alpha_{n,c}^{E2(k)}$&&&       6.9  &  0.3 \\
$\alpha_{n,cv}^{E2(k)}$&&&     0  &  0 \\
& & \\
Total($\alpha_{n}^{E2(k)}$)&& &          875  &       10 \\
& & \\
Others~\cite{safsaf} & & & 871 & 4 \\
Others~\cite{mitroy} & &  &875.1&   \\
%\end{longtable}
\end{tabular}
\begin{tabular}{lrrrrrr}
\multicolumn{7}{c}{3$D_{3/2}$ state} \\
\multicolumn{1}{c}{Contribution} &
\multicolumn{1}{c}{$Q$} &
\multicolumn{1}{c}{$\delta Q$} &
\multicolumn{1}{c}{$\alpha_{n}^{E2(0)}$} &
\multicolumn{1}{c}{$\delta\alpha_{n}^{E2(0)}$} &
\multicolumn{1}{c}{$\alpha_{n}^{E2(2)(0)}$} &
\multicolumn{1}{c}{$\delta\alpha_{n}^{E2(2)(0)}$}\\
\hline 
 & & \\
\multicolumn{5}{l}{Main valence correlation} \\
3$D_{3/2}$  -  4$S_{1/2}$   &  7.971  &  0.064 &        -102.16  &     1.64  &        102.16  &     1.64 \\
3$D_{3/2}$  -  5$S_{1/2}$   &  3.933  &  0.069 &           8.81  &     0.31  &         -8.81  &     0.31 \\
3$D_{3/2}$  -  6$S_{1/2}$   &  1.044  &  0.006 &           0.42  &     0.01  &         -0.42  &     0.01 \\
3$D_{3/2}$  -  7$S_{1/2}$   &  0.545  &  0.001 &           0.10  &    $\sim$0&         -0.10  &    $\sim$0\\
%3$D_{3/2}$  -  8$_{/2}$ -1  &  0.356  &  0.001 &           0.039  &     $\sim$0&         -0.039  &    $\sim$0\\
3$D_{3/2}$  -  4$D_{3/2}$   &  6.381  &  0.029 &          20.69  &     0.19  &          0.00  &     0.00 \\
3$D_{3/2}$  -  5$D_{3/2}$   &  1.983  &  0.006 &           1.46  &     0.01  &          0.00  &     0.00 \\
3$D_{3/2}$  -  6$D_{3/2}$   &  1.075  &  0.003 &           0.38  &   $\sim$0 &          0.00  &     0.00 \\
3$D_{3/2}$  -  3$D_{5/2}$   &  3.794  &  0.036 &        5205.49  &    98.79  &       3718.21  &    70.56 \\
3$D_{3/2}$  -  4$D_{5/2}$   &  4.171  &  0.018 &           8.84  &     0.08  &          6.31  &     0.05 \\
3$D_{3/2}$  -  5$D_{5/2}$   &  1.298  &  0.004 &           0.63  &     0.01  &          0.45  &    $\sim$0 \\
3$D_{3/2}$  -  6$D_{5/2}$   &  0.704  &  0.002 &           0.16  &   $\sim$0 &          0.12  &    $\sim$0 \\
3$D_{3/2}$  -  5$G_{7/2}$   &  4.522  &  0.094 &           6.96  &     0.29  &         -1.99  &     0.08 \\
3$D_{3/2}$  -  6$G_{7/2}$   &  3.630  &  0.066 &           4.14  &     0.15  &         -1.18  &     0.04 \\
3$D_{3/2}$  -  7$G_{7/2}$   &  2.884  &  0.048 &           2.50  &     0.08  &         -0.71  &     0.02 \\
& & \\
Tail($\alpha_{n,v}^{E2(k)}$)  &  &    &       60 &      30  &          -17  & 8 \\
$\alpha_{n,c}^{E2(k)}$        &  &    &            6.9    &      0.3 & 0 & 0\\
$\alpha_{n,cv}^{E2(k)}$        &  &    &            0    &      0 & 0 & 0\\
& & \\
Total($\alpha_{n}^{E2(k)}$)   &  &    &  5225  &     103  &         $3797$  & 71 \\
\end{tabular}
\end{ruledtabular}
\end{table}

\begin{table*}[h!]
\caption{Contributions to different components of static quadrupole polarizability (in a.u.) of the $3D_{5/2}$ state of Ca$^+$. The absolute values 
and uncertainties (denoted by $\delta$) are for the corresponding contributions are also given.}\label{nd5ca}
\begin{ruledtabular}
\begin{tabular}{lrrrrrrrr}
\multicolumn{1}{c}{Contribution} &
\multicolumn{1}{c}{$Q$} &
\multicolumn{1}{c}{$\delta Q$} &
\multicolumn{1}{c}{$\alpha_{n}^{E2(0)}$} &
\multicolumn{1}{c}{$\delta\alpha_{n}^{E2(0)}$} &
\multicolumn{1}{c}{$\alpha_{n}^{E2(2)}$} &
\multicolumn{1}{c}{$\delta\alpha_{n}^{E2(2)}$} &
\multicolumn{1}{c}{$\alpha_{n}^{E2(4)}$} &
\multicolumn{1}{c}{$\delta\alpha_{n}^{E2(4)}$}\\
\hline \\
\multicolumn{9}{l}{Main valence correlation} \\
3$D_{5/2}$  -  4$S_{1/2}$   &  9.785  &  0.081 &        -102.18  &     1.69  &        145.97  &     2.42  &        -21.90  &     0.36 \\
3$D_{5/2}$  -  5$S_{1/2}$   &  4.841  &  0.089 &           8.92  &     0.33  &        -12.74  &     0.47  &          1.91  &     0.07 \\
3$D_{5/2}$  -  6$S_{1/2}$   &  1.282  &  0.007 &           0.42  &     0.01  &         -0.60  &     0.01  &          0.09  &    $\sim$0  \\
3$D_{5/2}$  -  7$S_{1/2}$   &  0.670  &  0.002 &           0.10  &    $\sim$0 &         -0.14  &  $\sim$0  &          0.02  &    $\sim$0\\
3$D_{5/2}$  -  8$S_{1/2}$   &  0.437  &  0.001 &           0.04  &     $\sim$0&         -0.06  &     $\sim$0&          0.01  &    $\sim$0\\
3$D_{5/2}$  -  3$D_{3/2}$   &  3.794  &  0.036 &       -3470.33  &    65.86  &       1770.58  &    33.60  &        849.88  &    16.13 \\
3$D_{5/2}$  -  4$D_{3/2}$   &  4.194  &  0.022 &           5.98  &     0.06  &         -3.05  &     0.03  &         -1.46  &     0.02 \\
3$D_{5/2}$  -  5$D_{3/2}$   &  1.301  &  0.004 &           0.42  &   $\sim$0 &         -0.21  &  $\sim$0  &         -0.10  &     $\sim$0  \\
3$D_{5/2}$  -  6$D_{3/2}$   &  0.705  &  0.002 &           0.11  &   $\sim$0 &         -0.06  &     $\sim$0&         -0.03  &    $\sim$0\\
3$D_{5/2}$  -  4$D_{5/2}$   &  8.376  &  0.042 &          23.79  &     0.24  &         12.14  &     0.12  &          3.28  &     0.03 \\
3$D_{5/2}$  -  5$D_{5/2}$   &  2.602  &  0.008 &           1.68  &     0.01  &          0.86  &     0.01  &          0.23  &     $\sim$0  \\
3$D_{5/2}$  -  6$D_{5/2}$   &  1.411  &  0.005 &           0.44  &    $\sim$0&          0.22  &    $\sim$0&          0.06  &    $\sim$0\\
3$D_{5/2}$  -  5$G_{7/2}$   &  1.514  &  0.033 &           0.52  &     0.02  &          0.45  &     0.02  &         -0.02  &    $\sim$0  \\
3$D_{5/2}$  -  6$G_{7/2}$   &  1.215  &  0.023 &           0.31  &     0.01  &          0.27  &     0.01  &         -0.01  &    $\sim$0\\
3$D_{5/2}$  -  7$G_{7/2}$   &  0.965  &  0.016 &           0.19  &     0.01  &          0.16  &     0.01  &         -0.01  &    $\sim$0\\
3$D_{5/2}$  -  5$G_{9/2}$   &  5.353  &  0.115 &           6.51  &     0.28  &         -3.32  &     0.14  &          0.03  &     $\sim$0  \\
3$D_{5/2}$  -  6$G_{9/2}$   &  4.295  &  0.080 &           3.87  &     0.14  &         -1.97  &     0.07  &          0.02  &     $\sim$0  \\
3$D_{5/2}$  -  7$G_{9/2}$   &  3.413  &  0.059 &           2.33  &     0.08  &         -1.19  &     0.04  &          0.01  &    $\sim$0\\
& & \\
Tail($\alpha_{n,v}^{E2(k)}$)&&&         60  &      30  &          -24 &     12 &           0.2  &      0.1 \\
$\alpha_{n,c}^{E2(k)}$&&&     6.9  &  0.3 & 0 & 0 & 0 & 0\\
$\alpha_{n,cv}^{E2(k)}$&&&     0  &  0 & 0 & 0 & 0 & 0\\
& & \\
Total($\alpha_{n}^{E2(k)}$) &  &&    $-3450$  &      72  &         $1883$  &      36  &       $831$  &     16 \\ 
\end{tabular}
\end{ruledtabular}
\end{table*}

\section{Tensor components of $\alpha^{E2}$}

The time-dependent interaction Hamiltonian $H_{int}$ describing the interaction between an 
atomic system and the gradient of electric field generated by laser is expressed as
\begin{equation}
H_{int}(r,t)=-{\bf \nabla {\cal E}}({\bf r},t)\cdot{\bf Q},
\end{equation}
where {\bf Q} is the E2 operator and ${\bf \nabla {\cal E}} (t)$ is a tensor representing the gradient of 
time-dependent electric field ${\cal E}({\bf r},t)$ written as ${\cal E}({\bf r},t)= \frac{1}{2}|{\cal E}(r)| \hat{\epsilon} e^{-\iota \omega t
}+ c.c.$ with $\hat{\epsilon}$ and $\omega$ representing the polarization vector and angular frequency respectively. The 
above interaction Hamiltonian is Harmonic in time and  varies periodically as
\begin{equation}
H_{int}(r,t) =V_1(r) e^{-\iota\omega t}+V_2(r)e^{\iota\omega t},
\end{equation}
with $V_1(r)= -\frac{{\bf \nabla \cal E}\cdot{\bf Q}}{2}$, $V_2(r)=-\frac{{\bf \nabla \cal E^*}\cdot{\bf Q}}{2}$ and $\omega$ is the frequency of the 
oscillating electric field. For the periodic oscillating electric field, the time-dependent wave functions of the 
wave function of $n^{th}$ state can be expressed as 
\begin{eqnarray}
 | n,r,t \rangle  = |n, r \rangle e^{ip \omega t},
\end{eqnarray}
where $p=0, \pm 1, \pm 2, \cdots$ to satisfy the periodicity requirement. By adopting Floquet perturbation expansion approach 
\cite{sambe}, it can be shown that 
\begin{eqnarray}
\langle \langle n',r',t| V_{1,2} e^{\pm i p \omega t} | n,r,t \rangle \rangle &=& \langle n',r| V_{1,2} | n,r \rangle \delta_{p,p\pm 1} .
\end{eqnarray}
Therefore, using the angular momentum notation of the state as $|\gamma _n J_n M_{n}\rangle$ with angular momentum $J_n$, magnetic 
projection $M_{n}$ and $\gamma_n$ representing other quantum numbers, the second-order change in energy can be expressed as
\begin{equation}
\Delta E_n^{E2(2)}=\langle\gamma_n J_n M_n|H_{eff}|\gamma_n J_n M_n \rangle,
\label{qdelta}
\end{equation}
where the time-independent effective Hamiltonian $H_{eff}$ is defined as
\begin{equation}\nonumber
H_{eff}(r) =V_1(r) {\cal R}_n^-V_2(r)+V_2(r) {\cal R}_n^+V_1(r),
\end{equation}
with the projection operators $R_n^{\pm}$ given by ~\cite{Beloy}
\begin{eqnarray}
 {\cal R}_n^{\pm} = \sum_{m \ne n} \frac{|\gamma_m J_m M_m \rangle\langle\gamma_m J_m M_m|}{E_n-E_m\pm\omega} .
\end{eqnarray}
Substituting the respective expressions for $V_1$ and $V_2$, it yields
\begin{eqnarray}\nonumber
H_{eff}&=&\left(\frac{1}{2}\right)^2 \left[ \left({\bf \nabla \cal E}\cdot{\bf Q}\right) {\cal R}_n^-\left({\bf \nabla \cal E^*}\cdot{\bf Q}\right)\right.\\
&& + \left.\left({\bf \nabla \cal E^*}\cdot{\bf Q}\right) {\cal R}_n^+\left({\bf \nabla \cal E}\cdot{\bf Q}\right) \right] \nonumber \\
 &=&{\left(\frac{1}{2}\right)}^2\sum_{k=0}^4 \sum_{q=-k}^k (-1)^k \nonumber \\ && \times \left[
 \{{\bf \nabla \cal E^*}\otimes {\bf \nabla \cal E}\}^k_{q} 
 \left[\{{\bf Q}\otimes {\cal R}^+_n{\bf Q}\}^k_{q} \right. 
 \right.\nonumber\\
 && +\left.\left. \{{\bf Q}\otimes {\cal R}_n^-{\bf Q}\}^k_{q} \right]\right],
\label{qeqheff}
\end{eqnarray}
where $\{ \cdots \}$ represents the spherical tensor of rank $k$ with the allowed values for $q$ lying between $-k$ to $k$. The allowed values of $k$ are limited by the triangular selection rule. From the coupling of two of the rank-2 tensors, the allowed values of $k$ are restricted to  $0\le k \le 4$. 

This leads to the energy shift formula as
\begin{eqnarray}\nonumber
\Delta E_n^{E2(2)}&=&{\left(\frac{1}{2}\right)}^2\sum_k \sum_{q=-k}^k (-1)^{2J_n} \{{\bf \nabla \cal E^*}
\otimes{\bf \nabla \cal E}\}^k_{q} \nonumber \\
&& (-1)^{J_n-M_{n}} \left( \begin{array}{ccc}
                                            J_n& k & J_n\\
                                            -M_{n} & q & M_{n} 
                                           \end{array}\right)
                                           (2k+1)^{1/2} \nonumber \\
&& \times \sum_{m \ne n } (-1)^{J_n-J_m} \left\{ \begin{array}{ccc}
                                            J_n& k & J_n\\
                                            2 & J_m & 2 
                                           \end{array}\right\} \nonumber \\
                           && \times|\langle \gamma_n J_n||{\bf Q}||\gamma_m J_m\rangle|^2 \nonumber \\
                           && \times \left [ \frac{1}{E_n-E_m+\omega} 
       +(-1)^k \frac{1}{E_n-E_m-\omega} \right ] .\nonumber\\
\end{eqnarray}
Only $k=0$ component is allowed in the above expression due to the selection rule associated with the 3-j symbol $\left( \begin{array}{c} \cdots \\ \cdots \end{array} \right) $. 

For simplification we assume applied external electric field as linearly polarized
 and its gradient only in the z-direction. In this case, the allowed values for $k$ are restricted to 0, 2 and 4 and we have \cite{wmitano}
\begin{eqnarray}\nonumber
\{{\bf \nabla \cal E^*}\otimes{\bf \nabla \cal E}\}^0_{0}&=&\frac{1}{4\sqrt{5}}\left(\frac{\partial E_z}{\partial z}\right)^2,\\ \nonumber
\{{\bf \nabla \cal E^*}\otimes{\bf \nabla \cal E}\}^2_{0}&=&-\frac{1}{4}\sqrt{\frac{2}{7}}\left(\frac{\partial E_z}{\partial z}\right)^2,
\end{eqnarray}
and
\begin{eqnarray} \nonumber
\{{\bf \nabla \cal E^*}\otimes{\bf \nabla \cal E}\}^4_{0}&=&\frac{3}{4}\sqrt{\frac{2}{35}}\left(\frac{\partial E_z}{\partial z}\right)^2.
\end{eqnarray}
 Using these factors, we can divide the total second-order energy shift into  
\begin{equation}
\Delta E_n^{E2(2)}= \sum_{k=0,2,4} \Delta E_{n}^{(2,k)}(\omega),
\label{eshi}
\end{equation}
where $\Delta E_{n}^{E2(2,0)}(\omega)$, $\Delta E_{n}^{E2(2,2)}(\omega)$ and $\Delta E_{n}^{E2(2,4)}(\omega)$ are the terms corresponding 
to $k=0$, 2 and 4, respectively. This follows the expression for $k=0$ as 
\begin{widetext}   
\begin{eqnarray}
\Delta E_{n}^{E2(2,0)}(\omega)&=&-\left(\frac{1}{16}\right)\left(\frac{\partial E_z}{\partial z}\right)^2\left(\frac{1}{5(2J_n+1)}\right)\sum_{m \neq n} \Big[\frac{|\langle \gamma_n J_n||{\bf Q}||\gamma_m J_m\rangle|^2}{E_m-E_n+\omega}
+\frac{|\langle \gamma_n J_n||{\bf Q}||\gamma_m J_m\rangle|^2}{E_m-E_n-\omega}\Big] .
\end{eqnarray}
Similarly, for $k=2$ is given by
\begin{eqnarray}
\Delta E_{n}^{E2(2,2)}(\omega)&=&-{\left(\frac{1}{16}\right)} \left(\frac{\partial E_z}{\partial z}\right)^2\frac{3M_{n}^2-J_n(J_n+1)}{J_n(2J_n-1)} \sqrt{\frac{10J_n(2J_n-1)}{7(J_n+1)(2J_n+1)(2J_n+3)}}\nonumber \\
&&\sum_{m \neq n}(-1)^{J_n+J_m+1}\left\{ \begin{array}{ccc}
                                            J_n& 2 & J_n\\
                                            2 & J_m & 2 
                                           \end{array}\right\}
                                         \left[\frac{|\langle \gamma_n J_n||{\bf Q}||\gamma_m J_m\rangle|^2}{E_m-E_n+\omega}\right.
                                       + \left.\frac{|\langle \gamma_n J_n||{\bf Q}||\gamma_m J_m\rangle|^2}{E_m-E_n-\omega}\right] 
\end{eqnarray}
 and for $k=4$ is given by
\begin{eqnarray}
\Delta E_{n}^{E2(2,4)}(\omega)&=& \left(\frac{1}{16}\right)\left(\frac{\partial E_z}{\partial z}\right)^2\left(9\sqrt{\frac{2}{35}}\right) \frac{[3(5M_n^2-J_n^2 -2J_n)(5M_n^2+1-J_n^2)-10M_n^2(4M_n^2-1)]}
                                           {\sqrt{(2J_n+3)(2J_n+2)(2J_n+1)(J_n)(2J_n-1)(J_n-1)(2J_n-3)}}\nonumber \\ 
                             &&  \sum_{m\neq n} \frac{(-1)^{J_n+J_m+1}}{\sqrt{(2J_n+5)(2J_n+4)}}\left\{ \begin{array}{ccc}
                                            J_n& 4 & J_n\\
                                            2 & J_m & 2 
                                           \end{array}\right\}
                       \left[\frac{|\langle \gamma_n J_n||{\bf Q}||\gamma_m J_m\rangle|^2}{E_m-E_n+\omega}\right. 
                                         +\left.\frac{|\langle \gamma_n J_n||{\bf Q}||\gamma_m J_m\rangle|^2}{E_m-E_n-\omega}\right].
\end{eqnarray}
\end{widetext}

The second-order energy can also be given in a compact form as
\begin{eqnarray}
 \Delta E_n^{E2(2)}(\omega)=-\frac{1}{16} \left(\frac{\partial E_z}{\partial z}\right)^2\alpha_{n}^{E2}(\omega) ,
\end{eqnarray}
where $\alpha_{n}^{E2}$ is known as the quadrupole polarizability of the state. We can conveniently evaluate $\alpha_{n}^{E2}$ 
by rewriting it in terms of different tensor components as
\begin{widetext}
\begin{eqnarray}
\alpha_{n}^{E2}(\omega)&=&\alpha_{n}^{E2(0)}(\omega)+\frac{3M_{n}^2-J_n(J_n+1)}{J_n(2J_n-1)}\alpha_{n}^{E2(2)}(\omega) + \frac{3(5M_n^2-J_n^2 -2J_n)(5M_n^2+1-J_n^2)-10M_n^2(4M_n^2-1)}{J_n(J_n-1)(2J_n-1)(2J_n-3)} \alpha_{n}^{E2(4)}(\omega),\nonumber \\
\end{eqnarray}
\end{widetext}
 where $\alpha_n^{E2(0)}(\omega)$, $\alpha_n^{E2(2)}(\omega)$ and $\alpha_n^{E2(4)}(\omega)$ are defined as the scalar, tensor component of rank 
2 and tensor component of rank 4 quadrupole dynamic polarizabilities, respectively. Collecting all the factors, the corresponding expressions 
for different components of quadrupole polarizability can be given by
\begin{eqnarray}
\alpha_{n}^{E2(0)}(\omega)&=& \sum_{m \ne n} W_{n}^{(0)} |\langle \gamma _nJ_n||{\bf Q}||\gamma _m J_m \rangle|^2 \nonumber \\
&& \times \left [\frac{ 1}{E_m -E_n +\omega}+\frac{ 1}{E_m-E_n-\omega}\right],\label{eqpolz1q} \\
 \alpha_{n}^{E2(2)}(\omega)&=& \sum_{m \ne n} W_{n,m}^{(2)} |\langle \gamma _nJ_n||{\bf Q}||\gamma _m J_m \rangle|^2 \nonumber \\ 
&& \times \left [\frac{ 1}{E_m -E_n +\omega}+\frac{1}{E_m-E_n-\omega}\right],\label{eqpolz2q}
\end{eqnarray}
and
\begin{eqnarray}
\alpha_{n}^{E2(4)}(\omega)&=& \sum_{m \ne n} W_{n,m}^{(4)} |\langle \gamma _nJ_n||{\bf Q}||\gamma_m J_m \rangle|^2 \nonumber \\
&& \times \left [\frac{1}{E_m -E_n +\omega}+\frac{ 1}{E_m-E_n-\omega}\right] \label{eqpolz3q}
\end{eqnarray}
with the coefficients
\begin{eqnarray}
W_{n}^{(0)} &=&\frac{1}{5(2J_n+1)}, \\
W_{n,m}^{(2)}&=&\sqrt{\frac{10J_n(2J_n-1)}{7(J_n+1)(2J_n+1)(2J_n+3)}}   \nonumber \\
              & & \times (-1)^{J_n+J_m+1}  \left\{ \begin{array}{ccc}
                             J_n& 2 & J_n\\
                          2 & J_m &2 
                         \end{array}\right\}, 
\end{eqnarray}
and
\begin{eqnarray}
W_{n,m}^{(4)} &=&-\sqrt{\frac{J_n(J_n-1)(2J_n-1)(2J_n-3)}{70(2J_n+1)(J_n+1)(J_n+2)}} \nonumber \\
  && \frac{9(-1)^{J_n+J_m+1}}{\sqrt{(2J_n+5)(2J_n+3)}} \left\{ \begin{array}{ccc}
                                            J_n& 4 & J_n\\
                                            2 & J_m & 2 
                                           \end{array}\right\}.  \ \ \ \ \ \                                       
\end{eqnarray}
It is obvious from the above expressions that only the scalar component will contribute to the states with angular momentum $J=1/2$, 
both scalar and tensor components with $k=2$ will contribute to the states with $J=3/2$ and all the components will contribute to 
the states with $J>3/2$. The above expressions can be used to determine both the static and dynamic quadrupole polarizabilities of the 
atomic states in a system, but we present here only the static values of these quantities.

\begin{table}[h!]
%\begin{ruledtabular}
%\begin{longtable}{lrrrr}
\caption{Contributions to the static scalar and rank 2 tensor component quadrupole polarizabilities (in a.u.) of the ground and $4D_{3/2}$ states of
Sr$^{+}$. The absolute values and uncertainties (denoted by $\delta$) are given from various contributions.}\label{nssr}
\begin{ruledtabular}
\begin{tabular}{lrrrr}
\multicolumn{5}{c}{Ground state} \\
\multicolumn{1}{c}{Contribution} &
\multicolumn{1}{c}{$Q$} &
\multicolumn{1}{c}{$\delta Q$} &
\multicolumn{1}{c}{$\alpha_{n}^{E2(0)}$} &
\multicolumn{1}{c}{$\delta \alpha_{n}^{E2(0)}$}\\
\hline
 & & \\
\multicolumn{5}{l}{Main valence correlation} \\
5$S_{1/2}$  -  4$D_{3/2}$    & 11.172  &  0.072 &         376.39  &     4.85 \\
5$S_{1/2}$  -  5$D_{3/2}$    & 12.959  &  0.028 &         138.34  &     0.60 \\
5$S_{1/2}$  -  6$D_{3/2}$    &  4.902  &  0.011 &          15.62  &     0.07 \\
5$S_{1/2}$  -  7$D_{3/2}$    &  2.789  &  0.010 &           4.58  &     0.03 \\
5$S_{1/2}$  -  4$D_{5/2}$    & 13.797  &  0.091 &         563.20  &     7.43 \\
5$S_{1/2}$  -  5$D_{5/2}$    & 15.757  &  0.037 &         204.19  &     0.96 \\
5$S_{1/2}$  -  6$D_{5/2}$    &  6.002  &  0.013 &          23.40  &     0.10 \\
5$S_{1/2}$  -  7$D_{5/2}$    &  3.425  &  0.013 &           6.90  &     0.05 \\
& & \\
Tail($\alpha_{n,v}^{E2(k)}$)&&&           30  &      15 \\
$\alpha_{n,c}^{E2(k)}$&&&      17.1  &  0.9 \\
$\alpha_{n,cv}^{E2(k)}$&&&      -0.0001  &  $\sim$0 \\
& & \\
Total($\alpha_n^{E2(k)}$)&& &         1379  &      17 \\
& & \\
Others~\cite{mitroy} & & &1346 & \\
%\end{longtable}
\end{tabular}
\begin{tabular}{lrrrrrr}
\multicolumn{7}{c}{4$D_{3/2}$ state} \\
\multicolumn{1}{c}{Contribution} &
\multicolumn{1}{c}{$Q$} &
\multicolumn{1}{c}{$\delta Q$} &
\multicolumn{1}{c}{$\alpha_{n}^{E2(0)}$} &
\multicolumn{1}{c}{$\delta\alpha_{n}^{E2(0)}$} &
\multicolumn{1}{c}{$\alpha_{n}^{E2(2)}$} &
\multicolumn{1}{c}{$\delta\alpha_{n}^{E2(2)}$}\\
\hline  
 & & \\
\multicolumn{5}{l}{Main valence correlation} \\
4$D_{3/2}$  -  5$S_{1/2}$   & 11.172  &  0.072 &        -188.20  &     2.43  &        188.20  &     2.43 \\
4$D_{3/2}$  -  6$S_{1/2}$   &  6.041  &  0.084 &          24.14  &     0.67  &        -24.14  &     0.67 \\
4$D_{3/2}$  -  7$S_{1/2}$   &  1.528  &  0.004 &           1.02  &     0.01  &         -1.02  &     0.01 \\
4$D_{3/2}$  -  8$S_{1/2}$   &  0.792  &  0.000 &           0.24  &     $\sim$0&         -0.23  &    $\sim$0\\
4$D_{3/2}$  -  5$D_{3/2}$   &  8.317  &  0.045 &          39.20  &     0.42  &          0.00  &     0.00 \\
4$D_{3/2}$  -  6$D_{3/2}$   &  2.701  &  0.007 &           3.02  &     0.02  &          0.00  &     0.00 \\
4$D_{3/2}$  -  7$D_{3/2}$   &  1.493  &  0.003 &           0.81  &  $\sim$0  &          0.00  &     0.00 \\
4$D_{3/2}$  -  4$D_{5/2}$   &  6.001  &  0.047 &        2819.33  &    44.16  &       2013.81  &    31.54 \\
4$D_{3/2}$  -  5$D_{5/2}$   &  5.401  &  0.028 &          16.49  &     0.17  &         11.78  &     0.12 \\
4$D_{3/2}$  -  6$D_{5/2}$   &  1.766  &  0.005 &           1.29  &     0.01  &          0.92  &     0.01 \\
4$D_{3/2}$  -  7$D_{5/2}$   &  0.978  &  0.002 &           0.35  &  $\sim$0  &          0.25  &    $\sim$0  \\
4$D_{3/2}$  -  5$G_{7/2}$   &  7.858  &  0.143 &          21.79  &     0.79  &         -6.23  &     0.23 \\
4$D_{3/2}$  -  6$G_{7/2}$   &  6.048  &  0.088 &          12.27  &     0.36  &         -3.51  &     0.10 \\
4$D_{3/2}$  -  7$G_{7/2}$   &  4.699  &  0.064 &           7.41  &     0.20  &         -2.12  &     0.06 \\
& & \\
Tail($\alpha_{n,v}^{E2(k)}$)&&&          108  &      54  &          -30  &     15 \\
$\alpha_{n,c}^{E2(k)}$&&&       17.1  &  0.7  & 0 & 0\\
$\alpha_{n,cv}^{E2(k)}$&&&       -0.0001 &  $\sim$0  & 0 & 0\\
& & \\
Total($\alpha_n^{E2(k)}$)&& &         2884  &      70  &         $2148$ &      35
\end{tabular}
\end{ruledtabular}
\end{table}

\begin{table*}[h!]
\caption{Contributions to different components of static quadrupole polarizability (in a.u.) of the $4D_{5/2}$ state of Sr$^+$. The absolute values 
and uncertainties (denoted by $\delta$) are for the corresponding contributions.}\label{nd5sr}
\begin{ruledtabular}
\begin{tabular}{lrrrrrrrr}
\multicolumn{1}{c}{Contribution} &
\multicolumn{1}{c}{$Q$} &
\multicolumn{1}{c}{$\delta Q$} &
\multicolumn{1}{c}{$\alpha_{n}^{E2(0)}$} &
\multicolumn{1}{c}{$\delta\alpha_{n}^{E2(0)}$} &
\multicolumn{1}{c}{$\alpha_{n}^{E2(2)}$} &
\multicolumn{1}{c}{$\delta\alpha_{n}^{E2(2)}$} &
\multicolumn{1}{c}{$\alpha_{n}^{E2(4)}$} &
\multicolumn{1}{c}{$\delta\alpha_{n}^{E2(4)}$}\\
\hline
\\
\multicolumn{9}{l}{Main valence correlation} \\
4$D_{5/2}$  -  5$S_{1/2}$   & 13.797  &  0.091 &        -187.73  &     2.48  &        268.19  &     3.54  &        -40.23  &     0.53 \\
4$D_{5/2}$  -  6$S_{1/2}$   &  7.550  &  0.109 &          25.35  &     0.73  &        -36.22  &     1.05  &          5.43  &     0.16 \\
4$D_{5/2}$  -  7$S_{1/2}$   &  1.894  &  0.005 &           1.05  &     0.01  &         -1.50  &     0.01  &          0.22  &     $\sim$0  \\
4$D_{5/2}$  -  8$S_{1/2}$   &  0.980  &  0.001 &           0.24  &     $\sim$0&         -0.34  & $\sim$0  &          0.05  &    $\sim$0\\
4$D_{5/2}$  -  4$D_{3/2}$   &  6.001  &  0.047 &       -1879.56  &    29.44  &        958.96  &    15.02  &        460.30  &     7.21 \\
4$D_{5/2}$  -  5$D_{3/2}$   &  5.538  &  0.033 &          11.67  &     0.14  &         -5.96  &     0.07  &         -2.86  &     0.03 \\
4$D_{5/2}$  -  6$D_{3/2}$   &  1.785  &  0.005 &           0.89  &     0.01  &         -0.45  &  $\sim$0  &         -0.22  &     $\sim$0  \\
4$D_{5/2}$  -  7$D_{3/2}$   &  0.984  &  0.002 &           0.24  &   $\sim$0 &         -0.12  &     $\sim$0&         -0.06  &    $\sim$0\\
4$D_{5/2}$  -  5$D_{5/2}$   & 10.991  &  0.063 &          45.87  &     0.53  &         23.40  &     0.27  &          6.32  &     0.07 \\
4$D_{5/2}$  -  6$D_{5/2}$   &  3.567  &  0.009 &           3.53  &     0.02  &          1.80  &     0.01  &          0.49  &     $\sim$0  \\
4$D_{5/2}$  -  7$D_{5/2}$   &  1.971  &  0.004 &           0.95  &   $\sim$0 &          0.49  &  $\sim$0  &          0.13  &    $\sim$0  \\
4$D_{5/2}$  -  5$G_{7/2}$   &  2.666  &  0.050 &           1.68  &     0.06  &          1.46  &     0.06  &         -0.07  &    $\sim$0  \\
4$D_{5/2}$  -  6$G_{7/2}$   &  2.048  &  0.030 &           0.94  &     0.03  &          0.82  &     0.02  &         -0.04  &     $\sim$0  \\
4$D_{5/2}$  -  7$G_{7/2}$   &  1.590  &  0.022 &           0.57  &     0.02  &          0.49  &     0.01  &         -0.02  &     $\sim$0  \\
4$D_{5/2}$  -  5$G_{9/2}$   &  9.425  &  0.175 &          21.00  &     0.78  &        -10.71  &     0.40  &          0.11  &     $\sim$0  \\
4$D_{5/2}$  -  6$G_{9/2}$   &  7.242  &  0.108 &          11.78  &     0.35  &         -6.01  &     0.18  &          0.06  &     $\sim$0  \\
4$D_{5/2}$  -  7$G_{9/2}$   &  5.621  &  0.078 &           7.10  &     0.20  &         -3.62  &     0.10  &          0.04  &     $\sim$0  \\

\\
Tail($\alpha_{n,v}^{E2(k)}$)&&&        110  &      55  &          -43  &     -22  &          0.4  &      0.2 \\
$\alpha_{n,c}^{E2(k)}$&&&     17.1  &  0.7 & 0 & 0 & 0 & 0 \\
$\alpha_{n,cv}^{E2(k)}$&&&     -0.0001  &  $\sim$0& 0 & 0 & 0 & 0 \\
\\
Total($\alpha_{n}^{E2(k)}$)&& &      -1807  &      62  &         1147  &      27  &       430  &     7
\end{tabular}
\end{ruledtabular}
\end{table*}

\begin{table}[h!]
%\begin{ruledtabular}
%\begin{longtable}{lrrrr}
\caption{Contributions to the static scalar and rank 2 tensor component quadrupole polarizabilities (in a.u.) the ground and $5D_{3/2}$ states of
Ba$^{+}$. The absolute values and uncertainties (denoted by $\delta$) are given from various contributions.}\label{nsba}
\begin{ruledtabular}
\begin{tabular}{lrrrr}
\multicolumn{5}{c}{Ground state} \\
\multicolumn{1}{c}{Contribution} &
\multicolumn{1}{c}{$Q$} &
\multicolumn{1}{c}{$\delta Q$} &
\multicolumn{1}{c}{$\alpha_{n}^{E2(0)}$} &
\multicolumn{1}{c}{$\delta \alpha_{n}^{E2(0)}$}\\
\hline
\\
\multicolumn{5}{l}{Main valence correlation} \\ 
6$S_{1/2}$  -  5$D_{3/2}$   & 12.740  &  0.151 &        1461.78  &    34.65 \\
6$S_{1/2}$  -  6$D_{3/2}$   & 16.797  &  0.018 &         269.52  &     0.58 \\
6$S_{1/2}$  -  7$D_{3/2}$   &  5.734  &  0.030 &          24.13  &     0.25 \\
6$S_{1/2}$  -  8$D_{3/2}$   &  3.166  &  0.023 &           6.60  &     0.10 \\
6$S_{1/2}$  -  5$D_{5/2}$   & 15.943  &  0.185 &        1966.09  &    45.63 \\
6$S_{1/2}$  -  6$D_{5/2}$   & 20.259  &  0.019 &         390.33  &     0.73 \\
6$S_{1/2}$  -  7$D_{5/2}$   &  7.042  &  0.035 &          36.34  &     0.36 \\
6$S_{1/2}$  -  8$D_{5/2}$   &  3.915  &  0.028 &          10.08  &     0.14 \\
\\
Tail($\alpha_{n,v}^{E2}(k)$)&&&           40  &      20 \\
$\alpha_{n,c}^{E2(k)}$&&&       46.0  &  2.3 \\
$\alpha_{n,cv}^{E2(k)}$&&&     -0.001 & $\sim$0 \\
\\
Total($\alpha_{n}^{E2(k)}$)&& &         4251  &      61 \\
\\
Others~\cite{bijaya022} & & &4270 & 27 \\
Others~\cite{Iskrenova} & & &4182 &34 \\
Others~\cite{uisaf} & & & 4091.5& \\
Others~\cite{tang} & & & 4821 & \\
Expt.~\cite{snow} & & &4420 & 250
%\end{longtable}
\end{tabular}
\begin{tabular}{lrrrrrr}
\multicolumn{7}{c}{5$D_{3/2}$ state} \\
\multicolumn{1}{c}{Contribution} &
\multicolumn{1}{c}{$Q$} &
\multicolumn{1}{c}{$\delta Q$} &
\multicolumn{1}{c}{$\alpha_{n}^{E2(0)}$} &
\multicolumn{1}{c}{$\delta\alpha_{n}^{E2(0)}$} &
\multicolumn{1}{c}{$\alpha_{n}^{E2(2)}$} &
\multicolumn{1}{c}{$\delta\alpha_{n}^{E2(2)}$}\\
\hline
\\
\multicolumn{5}{l}{Main valence correlation} \\
5$D_{3/2}$  -  6$S_{1/2}$   & 12.740  &  0.151 &        -730.89  &    17.33  &        730.89  &    17.33 \\
5$D_{3/2}$  -  7$S_{1/2}$   &  4.806  &  0.156 &          13.53  &     0.88  &        -13.53  &     0.88 \\
5$D_{3/2}$  -  8$S_{1/2}$   &  1.498  &  0.018 &           0.93  &     0.02  &         -0.93  &     0.02 \\
5$D_{3/2}$  -  9$S_{1/2}$   &  0.821  &  0.005 &           0.24  &   $\sim$0 &         -0.24  &   $\sim$0  \\
5$D_{3/2}$  -  6$D_{3/2}$   &  8.338  &  0.132 &          37.15  &     1.18  &          0.00  &     0.00 \\
5$D_{3/2}$  -  7$D_{3/2}$   &  2.987  &  0.017 &           3.57  &     0.04  &          0.00  &     0.00 \\
5$D_{3/2}$  -  8$D_{3/2}$   &  1.720  &  0.007 &           1.05  &     0.01  &          0.00  &     0.00 \\
5$D_{3/2}$  -  5$D_{5/2}$   &  6.772  &  0.086 &        1256.64  &    31.92  &        897.60  &    22.80 \\
5$D_{3/2}$  -  6$D_{5/2}$   &  5.338  &  0.085 &          15.15  &     0.48  &         10.82  &     0.35 \\
5$D_{3/2}$  -  7$D_{5/2}$   &  1.938  &  0.012 &           1.50  &     0.02  &          1.07  &     0.01 \\
5$D_{3/2}$  -  8$D_{5/2}$   &  1.120  &  0.005 &           0.45  &  $\sim$0  &          0.32  &   $\sim$0  \\
5$D_{3/2}$  -  5$G_{7/2}$   &  8.559  &  0.259 &          27.65  &     1.67  &         -7.90  &     0.48 \\
5$D_{3/2}$  -  6$G_{7/2}$   &  6.612  &  0.164 &          15.10  &     0.75  &         -4.31  &     0.21 \\
5$D_{3/2}$  -  7$G_{7/2}$   &  5.145  &  0.113 &           8.70  &     0.38  &         -2.49  &     0.11 \\
\\
Tail($\alpha_{n,v}^{E2(k)}$)&&&        123  &      62  &          -33  &     17 \\
$\alpha_{n,c}^{E2(k)}$&&&     46.0  &  2.3 & 0 & 0\\
$\alpha_{n,cv}^{E2(k)}$&&&     -0.001  &  $\sim$0 & 0 & 0\\
Total($\alpha_{n}^{E2(k)}$)&& &        820  &      72  &         1579 &      33 \\
\\
Others~\cite{bijaya022} & & & 835& 32
\end{tabular}
\end{ruledtabular}
\end{table}

\section{Evaluation of wave functions}

We use an all-order perturbative method in the relativistic coupled-cluster (RCC) theory framework with the singles and doubles excitations 
(SD method) approximation \cite{safronova12,safronova22} to determine the wave functions of the ground and metastable states of the considered 
alkaline earth-metal ions as described in our earlier works \cite{jasmeet}. In this approach, we obtain first the Dirac-Hartree-Fock (DHF) wave 
function ($|\Phi_0\rangle$) for the closed core of the alkali atomic configurations of the corresponding alkaline earth-metal ions. The DHF 
wave functions of the required states are then obtained by appending the respective valence orbital ($v$) with $|\Phi_0\rangle$ as 
$\vert \Phi_v \rangle= a_v^{\dagger}|\Phi_0\rangle$. The exact states using these DHF wave functions are expressed in the SD approximation as 
\begin{eqnarray}{\label{wav}}
|\Psi_v \rangle &\simeq& \{1+T+S_v\}|\Phi_v\rangle , 
\end{eqnarray}
where the operators $T$ and $S_v$ are responsible for the excitations of the core and valence electrons, respectively, from the DHF 
wave functions due to the correlation effects. In the second quantization notation, they are given as
\begin{eqnarray}
 T= \sum_{a,p} \rho_a^p a_p^{\dagger} a_a + \frac{1}{4} \sum_{a,b,p,q}\rho_{ab}^{pq} a_p^{\dagger} a_q^{\dagger} a_b a_a  
\end{eqnarray}
and 
\begin{eqnarray}
 S_v= \sum_{p\ne v} \rho_v^p a_p^{\dagger} a_v + \frac{1}{2} \sum_{p,q,a} \rho_{va}^{pq} a_p^{\dagger} a_q^{\dagger} a_a a_v  ,
\end{eqnarray}
where $a,b$ and $p,q$ electrons are from the occupied and unoccupied orbitals, respectively, and $\rho$ are the corresponding excitation 
amplitudes. To improve the results further, we also include contributions from the important triple excitations perturbatively in the 
determination of amplitudes of the excitation operators of the SD method (SDpT method) as defined in Refs. \cite{safronova12,safronova22}.

As demonstrated in Refs. \cite{nandy,jasmeet}, contributions to the square of the matrix element $|\langle \Psi_v | Q | \Psi_m \rangle|^2$
using the aforementioned procedure can be expressed as
\begin{eqnarray}
 \sum_m |\langle \Psi_v | Q | \Psi_m \rangle |^2 = Q_c +Q_{cv} + Q_v ,
\end{eqnarray}
where $Q_c$, $Q_{cv}$ and $Q_v$ denote correlation contributions due to the core, core-valence and valence electrons, respectively. 
Segregating these contributions in Eqs. (\ref{eqpolz1q}), (\ref{eqpolz2q}) and (\ref{eqpolz3q}), we can express 
\begin{eqnarray}
\alpha_{n}^{E2(k)}  &=& \alpha_{n,c}^{E2(k)} + \alpha_{n,cv}^{E2(k)} + \alpha_{n,v}^{E2(k)}
\label{eq26}
\end{eqnarray}
for the components $k=0,2$ and 4. It can be shown that core contributions vanish for $j=2$ and 4. The most important contributions  to different components of quadrupole polarizability arises 
through $\alpha_{n,v}^{E2(k)}$ since it takes into account the dominant correlation effects. 
We use the sum-over-states approach to estimate this contribution from many low-lying states by evaluating the electric quadrupole (E2) matrix 
elements in the SD method as
\begin{eqnarray}
\langle \Psi_v| Q| \Psi_m\rangle &=& \frac{\langle\Phi_v| \{ 1+T^{\dagger} + S_v^{\dagger} \} Q \{1+T+S_m\} |\Phi_m\rangle}{\sqrt{{\cal N}_v {\cal N}_m}} , \nonumber\\
\end{eqnarray}
where the normalization factor for the wave functions of the states with the respective valence orbital $v$ is given by ${\cal N}_v=
\{1+ T^{\dagger} T+ S_v^{\dagger} S_v \}$. These values with the experimental excitation energies are used to reduce the uncertainties in 
the evaluation of $\alpha_{n,v}^{E2(k)}$ contributions. Contributions from the higher excited states including continuum to the static 
values of $\alpha_{n}^{E2(k)}(v)$ are estimated using the expression
\begin{equation}
\alpha_{n,v}^{E2(k)}=-16\langle\Psi_n|Q|\Psi_n^{(1)}\rangle/\left(\frac{\partial E_z}{\partial z}\right)^2,
\end{equation}
where $|\Psi_n^{(1)}\rangle$ is the first-order correction to $| \Psi_n\rangle$ with energy $E_n$ due to the quadrupole operator $Q$. Using the 
inhomogeneous equation, we obtain its contribution at the DHF approximation of the Dirac-Coulomb atomic Hamiltonian ($H_{at}$) as
\begin{equation}
(H_{at}-E_n)|\Psi_n^{(1)}\rangle= (E_n^{E2(1)}-H_{eff})|\Psi_n^{(0)}\rangle .
\end{equation}
In the above expression, $E_n^{E2(1)}$ corresponds to the first-order change in energy due to the respective component of $H_{eff}$. 
By subtracting the DHF contribution due to the orbitals with principal quantum numbers corresponding to the states included in the sum-over-states approach, we 
extract out the contributions from the higher level excitations and denote it as ``Tail($\alpha_{n,v}^{E2(k)}$)''. Similarly, we 
use the DHF wave functions to determine the core-valence correlation contributions as they are found to be negligibly small. The 
core correlation contributions to the scalar component of the quadrupole polarizability values are obtained using random phase 
approximation (RPA) in the considered ions. A very large basis is used in all calculations carried out in this work. We use 70 
basis set functions for all partial waves with maximum number of orbital quantum number $l_{\rm max} \le$ 6.
\begin{table*}[h!]
\caption{Contributions to different components of static quadrupole polarizability (in a.u.) of the $5D_{5/2}$ state of Ba$^+$. The absolute values 
and uncertainties (denoted by $\delta$) are for the corresponding contributions.}\label{nd5ba}
\begin{ruledtabular}
\begin{tabular}{lrrrrrrrr}
\multicolumn{1}{c}{Contribution} &
\multicolumn{1}{c}{$Q$} &
\multicolumn{1}{c}{$\delta Q$} &
\multicolumn{1}{c}{$\alpha_{n}^{E2(0)}$} &
\multicolumn{1}{c}{$\delta\alpha_{n}^{E2(0)}$} &
\multicolumn{1}{c}{$\alpha_{n}^{E2(2)}$} &
\multicolumn{1}{c}{$\delta\alpha_{n}^{E2(2)}$} &
\multicolumn{1}{c}{$\alpha_{n}^{E2(4)}$} &
\multicolumn{1}{c}{$\delta\alpha_{n}^{E2(4)}$}\\
\hline
\\
\multicolumn{9}{l}{Main valence correlation} \\
5$D_{5/2}$  -  6$S_{1/2}$   & 15.943  &  0.185 &         -655.36  &     15.21  &         936.23  &     21.73  &        -140.43  &      3.26 \\
5$D_{5/2}$  -  7$S_{1/2}$   &  6.248  &  0.192 &           15.57  &      0.96  &         -22.25  &      1.37  &           3.34  &      0.21 \\
5$D_{5/2}$  -  8$S_{1/2}$   &  1.915  &  0.020 &            1.02  &      0.02  &          -1.46  &      0.03  &           0.22  &      $\sim$0  \\
5$D_{5/2}$  -  9$S_{1/2}$   &  1.045  &  0.006 &            0.27  &    $\sim$0 &          -0.38  &    $\sim$0 &           0.06  &    $\sim$0  \\
5$D_{5/2}$  -  5$D_{3/2}$   &  6.772  &  0.086 &         -837.76  &     21.28  &         427.43  &     10.86  &         205.17  &      5.21 \\
5$D_{5/2}$  -  6$D_{3/2}$   &  5.703  &  0.086 &           11.82  &      0.36  &          -6.03  &      0.18  &          -2.89  &      0.09 \\
5$D_{5/2}$  -  7$D_{3/2}$   &  2.009  &  0.011 &            1.09  &      0.01  &          -0.56  &      0.01  &          -0.27  &      $\sim$0  \\
5$D_{5/2}$  -  8$D_{3/2}$   &  1.151  &  0.005 &            0.32  &     $\sim$0&          -0.16  &   $\sim$0  &          -0.08  &      $\sim$0  \\
5$D_{5/2}$  -  6$D_{5/2}$   & 11.171  &  0.170 &           45.11  &      1.37  &          23.01  &      0.70  &           6.21  &      0.19 \\
5$D_{5/2}$  -  7$D_{5/2}$   &  3.991  &  0.023 &            4.30  &      0.05  &           2.19  &      0.03  &           0.59  &      0.01 \\
5$D_{5/2}$  -  8$D_{5/2}$   &  2.295  &  0.009 &            1.26  &      0.01  &           0.64  &      0.01  &           0.17  &      $\sim$0  \\
5$D_{5/2}$  -  5$G_{7/2}$   &  2.980  &  0.088 &            2.27  &      0.13  &           1.97  &      0.12  &          -0.09  &      0.01 \\
5$D_{5/2}$  -  6$G_{7/2}$   &  2.291  &  0.055 &            1.22  &      0.06  &           1.06  &      0.05  &          -0.05  &      $\sim$0  \\
5$D_{5/2}$  -  7$G_{7/2}$   &  1.778  &  0.037 &            0.70  &      0.03  &           0.61  &      0.03  &          -0.03  &      $\sim$0  \\
5$D_{5/2}$  -  5$G_{9/2}$   & 10.534  &  0.309 &           28.31  &      1.66  &         -14.44  &      0.85  &           0.14  &      0.01 \\
5$D_{5/2}$  -  6$G_{9/2}$   &  8.101  &  0.195 &           15.30  &      0.74  &          -7.81  &      0.38  &           0.08  &      $\sim$0  \\
5$D_{5/2}$  -  7$G_{9/2}$   &  6.288  &  0.134 &            8.76  &      0.37  &          -4.47  &      0.19  &           0.04  &      $\sim$0  \\
\\
Tail($\alpha_{n,v}^{E2(k)}$)&&&          128  &      64  &           -50  &      25  &           0.6  &      0.3 \\
$\alpha_{n,c}^{E2(k)}$&&&      46.0  &  2.3&0&0&0&0 \\
$\alpha_{n,cv}^{E2(k)}$&&&      -0.001  &  $\sim$0 &0&0&0&0\\
\\
Total($\alpha_{n}^{E2(k)}$)&& &        -1182  &      69  &         1286  &      35 &           72  &      6 \\
\\                                                                                                                
Others~\cite{bijaya022} & & &-1201 & 36
\end{tabular}
\end{ruledtabular}
\end{table*}

\section{Results and Discussion}

In Tables~\ref{nsca} and \ref{nd5ca}, we present the calculated scalar static quadrupole polarizabilities of the ground, $3D_{3/2}$ and $3D_{5/2}$ states of 
Ca$^+$. The $D$ state polarizabilities have contributions from the scalar as well as tensor components to the E2 polarizability. As shown in Table~\ref{nsca} 
the valence correlation contributions to the scalar and tensor quadrupole polarizabilities of the $4S$ and $3D_{3/2} $ states are the dominant ones. The core 
contribution calculated using the RPA approximation is found to be 6.9 in atomic units (a.u.). The uncertainty in this RPA value is expected to be on the order of 4\%. 
The $\alpha_{n,vc}^{E2(k)}$ contributions are estimated using the DHF method and are listed in the same table. The $\alpha_{n,v}^{E2(k)}$ contributions for the 
scalar and tensor part are evaluated using a sum-over-states approach. We use E2 matrix elements from the SDpT method and experimental energies from the NIST
database \cite{NIST} for the dominant transitions for these estimations. Contributions from these transitions to the quadrupole polarizabilities are listed 
explicitly in the above tables along with the respective values of the quadrupole matrix elements and corresponding uncertainties. The remainder 
Tail($\alpha_{n,v}^{E2(k)}$) contributions are evaluated in the DHF approximation. These contributions are found to be about $2\%$ and $1\%$ for the $4S$ 
and $3D_{3/2}$ states, respectively, of their total values. As can be seen, the E2 matrix elements of the $4S_{1/2}-3D_{3/2,5/2}$ and $4S_{1/2}-4D_{3/2,5/2}$ 
transitions have overwhelmingly dominant contributions to the scalar ground state quadrupole polarizability. Therefore, the uncertainty to the quadrupole 
polarizability of the ground state appear mainly from the errors to the E2 matrix elements of these transitions. Similarly, contribution from the E2 matrix 
element of the $3D_{3/2}-3D_{5/2}$ transition is preeminent for the scalar and tensor polarizabilities of the $3D_{3/2}$ state. Thus, it is important to consider error from 
this matrix element to estimate the uncertainty of the quadrupole polarizability of the $3D_{3/2}$ state. The uncertainties in the matrix elements are
determined as the difference between the final matrix elements calculated using the SDpT method and by scaling the calculations to account for the missing
correlation effects. Details of this scaling procedure can be found in Ref.~\cite{safronova}. Briefly, the single valence excitation coefficients are 
multiplied by the ratio of the corresponding experimental and theoretical correlation energies, and the matrix element calculation is repeated with the 
modified excitation coefficients. We assign maximum 50\% errors to the estimated tail contributions as the correlation effects are observed 
to contribute only about few percent in the evaluation of the E2 matrix elements in Ca$^+$. The net uncertainty to the quadrupole polarizabilities of the above 
states are given then by adding the individual uncertainties in quadrature. There are no experimental results available for these quantities, but we compare
our calculation for the ground state with two other calculations \cite{safsaf,mitroy}. Our value is 875(10) a.u. against that of 871(4) a.u. reported by 
Safronova and Safronova \cite{safsaf}. Mitroy and Zang~\cite{mitroy} have also obtained its value as 875.1 a.u. based on diagonalization of semi-empirical Hamiltonian 
in a large dimension single electron basis. Table~\ref{nd5ca} illustrates the scalar and tensor contributions for the E2 polarizability of the $3D_{5/2}$ state
of Ca$^+$ ion. It is evident from this table that an overwhelmingly dominant contribution to the scalar and tensor components of quadrupole polarizabilities
of the above state come from the E2 matrix element of the $3D_{5/2} - 3D_{3/2}$ transition. %Convergences in the results with the E2 matrix elements involving higher excited states are found to be faster than the ground state polarizability. Thus, the tail part is coming out to be very small. 

In Table~\ref{nssr}, we summarize the contribution to the static quadrupole polarizabilities of the ground $5S_{1/2}$ and the $4D_{3/2}$ states of the Sr$^+$ ion. 
The core contribution to the scalar quadrupole polarizability for Sr$^+$ is obtained from the RPA method as 17.1 a.u. We note that Tail($\alpha_{n,v}^{E2(k)}$) 
and $\alpha_{n,vc}^{E2(k)}$ contributions are realized to be small, therefore they are estimated in the DHF approximation. The Tail($\alpha_{n,v}^{E2(k)}$)
part contributes to $2\%$ and $4\%$ of the total valence polarizability for 5$S_{1/2}$ and 4$D_{3/2}$ states respectively and the uncertainty is taken to be 50\%. The first few dominant valence contributions 
calculated using the SDpT method along with the uncertainties have been explicitly shown in the table. We found one more calculation by Mitroy and Hang \cite{mitroy}
to compare our ground state quadrupole polarizability. Our calculations for the ground state of Sr$^+$ is 1379(17) a.u, which differs marginally by 2\% compared 
to the result reported in Ref.~\cite{mitroy}. Similar calculations for the $4D_{5/2}$ E2 polarizability for Sr$^+$ are presented in Table~\ref{nd5sr}. In this 
table, we present contributions from the scalar, rank 2 and rank 4 tensor contributions to the E2 polarizability of the above state explicitly. 

 The scalar quadrupole polarizability contribution to the $6S_{1/2}$ state and both the scalar and the tensor contributions to the E2 polarizabilities 
of the $5D_{3/2}$ state of Ba$^+$ have been illustrated in Table~\ref{nsba}. The core, core-valence and valence correlation contributions along with the 
individual dominant valence contributions from various low-lying states calculated using the sum-over-states approach are listed separately in the same table.
There have been several calculations of the ground state $6S_{1/2}$ quadrupole polarizabilities of this ion using different methods. Our ground state scalar
quadrupole polarizability value $4251$ a.u. is in agreement with the value of 4270 a.u. obtained by Sahoo and Das \cite{bijaya022}, which is carried out by 
including the non-linear terms of the RCC method in the singles and doubles excitations approximation. The analysis of Renovation-Tchaikovsky and 
Safronova \cite{Iskrenova} is based on relativistic all-order method like ours and found to be consistent with our result. Also, we find good agreement 
between our value with the calculations done by Safronova ~\cite{uisaf}, and Pail and Tang ~\cite{tang}. Unlike the previously discussed ions, there exists an experimental quadrupole polarizability value of $4420(25)$ a.u.
for the ground state  of the Ba$^+$ ion \cite{snow} and we find our result lies within the corresponding uncertainty of this 
measurement. The scalar quadrupole polarizabilities for the $5D_{3/2}$ and $5D_{5/2}$ states are found to be 820(72)a.u. and $-1182(69)$ a.u., which are in 
good agreement with the values of 835(32) a.u. and $-1201(36)$ a.u., respectively, from another calculation by Sahoo and Das ~\cite{bijaya022}. From this 
analysis, we believe that our tensor components to the quadrupole polarizabilities of the $5D_{3/2}$ and $5D_{5/2}$ states are quite reliable for their 
experimental verifications. 

\section{Conclusion}

We have given the scalar and tensor contributions to the electric quadrupole polarizabilities of the ground and metastable states of the singly charged 
calcium, strontium and barium ions.  Our calculations of the scalar polarizabilities are compared with the previously available theoretical and experimental 
results and they are found to be in good agreement. The derivations and the corresponding results for the tensor contributions to the quadrupole 
polarizabilities of the metastable states are given for their possible applications to high-precision experiments; especially for more accurate  estimates 
of systematic effects in the clock transitions of the above ions. Contributions from various correlations are given explicitly to these quantities. The 
dominant valence contributions are estimated by employing a relativistic all-order method and other smaller contributions are determined by lower-order 
relativistic many-body methods. Comparison of our results of the scalar polarizabilities with the previously available results suggests reliability of our tensor contributions to the quadrupole polarizabilities of the metastable states of the above ions. 
%The derivations of the formulas for the tensor components of the quadrupole polarizability can be used in other atomic systems for general purposes.

\section*{Acknowledgements}

The work of B.A. is supported by DST-SERB Grant No. EMR/2016/001228. The employed SDpT method was developed in the group of Professor M. S. Safronova of the 
University of Delaware, USA. B.K.S. would like to acknowledge the use of Vikram-100 HPC Supercomputer of Physical Research Laboratory, Ahmedabad, India.

%\bibliography{refsn.bib}
\end{document}